\documentclass[seceq]{ptptex}

\preprintnumber[3cm]{YGHP-07-40}

\title{BRST Symmetric Gaugeon Formalism for Higgs Model}

\author{
Hikaru \textsc{Miura}$^1$ and Ryusuke \textsc{Endo}$^{2,}$\footnote{%
E-mail: {endo@sci.kj.yamagata-u.ac.jp}}
}

\inst{%
$^1$Graduate School of Science and Engineering, Yamagata University\\ 
Yamagata 990-8560, Japan
\\
$^2$Department of Physics, Yamagata University, Yamagata 990-8560, Japan
}

\abst{%
We reinvestigate Yokoyama's gaugeon formalism for the spontaneously 
broken Abelian gauge theory.
Within the framework of the covariant linear gauges, 
we give a general gauge-fixing Lagrangian 
which includes the gauge field, the Goldstone mode, 
the multiplier $B$-field and Yokoyama's gaugeon fields 
(as well as Faddeev-Popov ghosts). 
As special choices of the values of the gauge-fixing parameters, 
our theory includes the usual covariant gauges and 
R$_\xi$-like gauges.
Although some of the gauge-fixing parameters can be 
shifted by the $q$-number gauge transformation, 
the $\xi$ parameter cannot be shifted in any of 
the R$_\xi$-like gauges.
}

\begin{document}

\maketitle

\section{Introduction}
In the standard formalism of canonically quantized gauge 
theories\cite{Nakanishi,K-O,Kugo}, 
we do not consider the gauge transformation on the quantum level.
There is no quantum gauge freedom,
since the quantum theory is defined only after the gauge fixing. 

Yokoyama's gaugeon formalism\cite{Y:QED,Y:book,Y:YM,Y:Higgs,Y:chiral,Y:Smatrix}
 provides a wider framework in which we can 
consider the quantum gauge transformation, 
$q$-number gauge transformation, among a family of Lorentz 
covariant linear gauges. 
In this formalism, a set of extra fields, so-called 
gaugeon fields, is introduced as the quantum gauge freedom. 
This formalism was proposed for
quantum electrodynamics\cite{Y:QED,Y:book}, 
spontaneously broken $U(1)$ gauge theory
\cite{Y:Higgs}, 
spontaneously broken chiral $U(1)$ gauge theory\cite{Y:chiral}  
and Yang-Mills gauge theory\cite{Y:YM}.  
Owing to the quantum gauge freedom, it becomes almost trivial to check the 
gauge parameter independence of the physical $S$-matrix\cite{Y:Smatrix}.

BRST symmetric theories of this formalism have been also developed 
for quantum electrodynamics\cite{Izawa,Koseki:QED,Endo}, 
Yang-Mills theory\cite{Abe,Koseki:YM} 
and the spin-3/2 Rarita-Schwinger gauge field\cite{Rarita}.
By virtue of the BRST symmetry, Yokoyama's physical 
subsidiary conditions have been improved\cite{Abe,Izawa,Koseki:QED}; 
the Gupta-Bleuler-type subsidiary conditions are replaced by a single 
Kugo-Ojima-type condition\cite{K-O,Kugo}. The BRST symmetry is also very
helpful to define the gauge invariant physical Hilbert 
space\cite{Koseki:QED,Endo,Rarita}.

In this paper we apply the BRST symmetric gaugeon formalism to
the Higgs model\cite{Higgs,Kibble}. 
The gaugeon formalism of this model was first studied 
by Yokoyama and Kubo\cite{Y:Higgs}. In their theory, 
the unphysical Goldstone boson mode appears as a 
massless dipole field; 
corresponding standard formalism is the theory 
of the usual covariant gauge\cite{Kibble,Nakanishi:Higgs,K-O}. 
Thus, 
their theory does not include other types of gauges, 
such as R$_\xi$ gauge\cite{Fujikawa,Yao}, where 
the Goldstone boson mode becomes massive.
The main purpose of the present paper is to construct a 
gaugeon formalism for the Higgs model which includes 
both the usual covariant gauges and R$_\xi$-like gauges.
In the R$_\xi$-like gauges, we also explore 
the possibility that the $\xi$ parameter might be 
shifted by the $q$-number gauge transformation.

In \S 2 we briefly review the Lorentz covariant quantization of 
the Higgs model; the theory in the usual covariant gauge (Lorenz gauge), 
the gaugeon formalism corresponding to this Lorenz gauge 
by Yokoyama and Kubo\cite{Y:Higgs}, and the theory of R$_\xi$-gauges 
by Fujikawa, Lee and Sanda and by Yao. 
In \S 3 we consider the most general gauge-fixing Lagrangian 
which consists of the gauge field, the Goldstone mode, the multiplier 
field and the gaugeon fields. The Lagrangian has seven gauge-fixing
parameters. As a special choice of the values of the parameters 
the theory would be equivalent ,for example, to the 
gaugeon formalism of the Lorenz gauge by Yokoyama and Kubo.  
By introducing two pairs of the Faddeev-Popov (FP) ghost fields to the
general Lagrangian given in \S 3, 
we present a general form of the BRST symmetric gaugeon formalism 
for the Higgs model in \S 4; the theory has the BRST symmetry, and 
admits the $q$-number gauge transformation under which some of 
the gauge-fixing parameters change their values. 
In \S 5 we study the R$_\xi$-like gauges of our theory. 
By choosing special values for the gauge-fixing 
parameters we show that our theory includes a 
 gaugeon formalism for the Yao's R$_\xi$ gauge, 
where one of two gauge-fixing parameter of Yao's 
theory can be shifted by the $q$-number gauge transformation, 
while the other ($\xi$-parameter) cannot be shifted. 
We also show that, in more general case, the $\xi$-parameter 
in any of the R$_\xi$-like gauges cannot be shifted 
by the $q$-number gauge transformation. 
\S 6 is devoted to a summary and discussion. The number of 
the conserved BRST-like charges in our theory is also discussed.

\section{Higgs model}
The Lagrangian of the Higgs model is given by
\begin{equation}
  \mathcal{L}_\mathrm{cl} 
     =
     - \frac{1}{4} F_{\mu \nu } F^{\mu \nu} 
     + (D_\mu \varphi )^\dagger (D^\mu \varphi ) 
     + \mu^2 \varphi^\dagger \varphi
     - \frac{\lambda}{2} (\varphi^\dagger \varphi )^2,
 \label{L:Higgs}
\end{equation}
where $F_{\mu\nu}=\partial_\mu A_\nu - \partial_\nu A_\mu$ 
is the field strength of the Abelian gauge field $A_\mu$, 
$\varphi$ is a complex scalar field, 
$D_\mu\varphi=(\partial_\mu - ieA_\mu)\varphi$ , 
$e$ is the charge of $\varphi$, and 
$\mu^2$ and $\lambda$ are positive constants. 
The vacuum expectation value of $\varphi$ is given by
$\langle 0|\varphi |0\rangle = v/\sqrt{2}= \sqrt{\mu^2/ \lambda}$, 
where we have adjusted the phase of $\varphi$ so that 
the vacuum expectation value is real.

The Lagrangian $\mathcal{L}_\mathrm{cl}$ is invariant 
under the gauge transformation,
\begin{align}
   A_\mu \ &\to \ A_\mu+ \partial _\mu \Lambda,
   \\
   \varphi(x) \ &\to \ e^{ie\Lambda(x)}\varphi(x). 
 \label{transf:gauge}
\end{align}
To quantize (\ref{L:Higgs}) we should choose a suitable gauge 
by adding a suitable gauge-fixing term (and a corresponding 
Faddeev-Popov(FP) ghost term) to $\mathcal {L}_\mathrm{cl}$.

\subsection{Lorenz gauge}
  The quantum Lagrangian in the usual covariant gauge (Lorenz gauge) 
is given by\cite{Kibble, Nakanishi:Higgs}
\begin{equation}
   \mathcal{L}_\mathrm{N} =   \mathcal{L}_\mathrm{cl}
             + B \partial_\mu  A^\mu + \frac{1}{2} \alpha B^2,
   \label{L:Lorenz}
\end{equation}
where $B$ is the multiplier $B$-field of 
Nakanishi-Lautrup\cite{Nakanishi,NL:B}, 
and the numerical constant $\alpha$ is the gauge-fixing parameter.
The BRST symmetric version is obtained by adding FP ghost 
term to (\ref{L:Lorenz}).\cite{K-O}\ 
The global $U(1)$ symmetry 
remains unbroken in this gauge, 
and the massless Goldstone boson arises as an unphysical mode. 
In particular, this massless Goldstone boson appears as a dipole 
field except for the Landau gauge $\alpha=0$. 

\subsection{gaugeon formalism for the Lorenz gauge}
By introducing the gaugeon fields $Y$ and $Y_*$ 
into (\ref{L:Lorenz}), we get the 
Lagrangian of Yokoyama and Kubo:\cite{Y:Higgs}
\begin{equation}
  \mathcal{L}_\mathrm{YK} = \mathcal{L}_\mathrm{cl}
             + B  \partial_\mu  A^\mu 
                + \frac{\varepsilon}{2}( a B + Y_*)^2
                - \partial_\mu Y_* \partial ^\mu Y,
   \label{L:YK}
\end{equation} 
where $\varepsilon$ is a sign factor ($\varepsilon=\pm 1$) and 
 $a$ is a numerical 
gauge-fixing parameter. The gauge-fixing parameter $\alpha$ in 
(\ref{L:Lorenz}) corresponds to $\varepsilon a^2$: 
the propagator%
   \footnote{%
     In this paper,  we use the symbol 
     $\langle \phi_A \phi_B \rangle$ to denote
     free propagators in momentum representation: 
     $
     \langle \phi_A \phi_B \rangle 
     = \mathrm {F.T.} 
                    \langle 0|T \phi_A(x) \phi_B(y)|0 \rangle,
    $
    in the interaction picture.
             } 
$\langle A_\mu A_\nu \rangle$ followed from (\ref{L:YK}) is 
exactly the same with that followed from (\ref{L:Lorenz}) if we assume
$\alpha=\varepsilon a^2$.

The Lagrangian (\ref{L:YK}) admits a $q$-number gauge transformation.
Under the field redefinition 
\begin{align}
  \hat A_\mu &= A_\mu + \tau \partial_\mu Y, 
  \qquad 
  \hat \varphi = e^{ie\tau Y}\varphi, 
  \label{q-transf:YK} \\
  \hat {Y_*} &= Y_* - \tau B, 
  \qquad
  \hat B = B, \qquad \hat Y=Y,
\end{align}
with $\tau$ being a numerical parameter, the Lagrangian is 
\textit{form invariant}, that is, it satisfies 
\begin{equation}
    \mathcal{L}_\mathrm{YK}(\phi_A, a) = 
    \mathcal{L}_\mathrm{YK}(\hat \phi_A, \hat a),
 \label{L:form_invariance}    
\end{equation}
where $\phi_A$ stands for any of the relevant fields and $\hat a$ 
is defined by
\begin{equation}
  \hat a = a+\tau.
\end{equation}
From the form invariance (\ref{L:form_invariance}), it can be immediately 
shown that  
the 
fields 
$\hat \phi_A$ and $\phi_A$ satisfy the same 
field equations and the same commutation relations 
except for the parameter $a$ 
which should be replaced by $\hat a$ for  $\hat \phi_A$. 

\subsection{R$_\xi$ gauge}
The quantum Lagrangian of R$_\xi$ gauge 
of Fujikawa, Lee and Sanda\cite{Fujikawa}
is given by 
\begin{equation}
   \mathcal{L}_\mathrm{FLS} =   \mathcal{L}_\mathrm{cl}
             - \frac{\xi}{2}\left( 
                            \partial_\mu  A^\mu 
                            + \frac{1}{\xi}M \chi 
                            \right)^2
   \label{L:FLS}
\end{equation}
in the Abelian case, where $\xi$ is a numerical gauge-fixing parameter, 
$M=ev$ denotes the acquired mass of $A_\mu$ through the spontaneous 
symmetry breaking, and the hermitian field $\chi$ is the Goldstone mode 
defined by
\begin{equation}
  \varphi = \frac{1}{\sqrt 2} (v + \psi + i\chi ),
 \label{phi:Re+Im}
\end{equation}
with $\psi$ being a physical Higgs mode. 
The global $U(1)$ symmetry is also broken through 
this gauge fixing so that the Goldstone mode $\chi$ acquires 
non-zero mass-squared  $M^2/\xi$. 
In particular, by taking the limit of $\xi \to 0$, 
we can reach the unitary gauge.
The BRST symmetric version of (\ref{L:FLS}) is  discussed 
in the textbook by Kugo\cite{Kugo}.

Yao has discussed a similar gauge.\cite{Yao}\  His Lagrangian can be 
read in our notation as 
\begin{equation}
   \mathcal{L}_\mathrm{Yao} =   \mathcal{L}_\mathrm{cl}
             +B \left( \partial_\mu  A^\mu +\frac{M}{\xi} \chi \right)
             + \frac{1}{2\eta}B^2,
   \label{L:Yao}
\end{equation}
where $\xi$ and $\eta$ are numerical gauge-fixing parameters. 
If we put $\xi=\eta$, the Lagrangian (\ref{L:Yao}) becomes identical with 
(\ref{L:FLS}), after eliminating the multiplier field $B$. 

Free propagators among $A_\mu$, $\chi$ and $B$ fields 
followed from (\ref{L:Yao}) are given by 
\begin{align}
 \langle A_\mu A_\nu \rangle 
       =& 
           \frac{1}{p^2 - M^2} \left(
                                      - g_{\mu\nu} + \frac{p_\mu p_\nu}{M^2}
                                \right)
         +\frac{p_\mu p_\nu}{M^2}
                \left[
                      - \frac{1}{p^2 -\xi^{-1}M^2}
                     + \frac{(\xi^{-1}-\eta^{-1})M^2}{{(p^2-\xi^{-1}M^2)}^2}
                \right],
\notag
\\
 \langle A_\mu \chi \rangle 
       =&\,
       - i(\xi^{-1}-\eta^{-1}) \frac{Mp_\mu}
             {(p^2 - \xi^{-1}M^2)^2},
\notag
\\
 \langle A_\mu B \rangle 
       =&\,
         \frac{ip_\mu}
             {p^2 - \xi^{-1}M^2},
\notag
 \\
 \langle \chi \chi \rangle 
        =& \,
          \frac{p^2-\eta^{-1}M^2}{(p^2-\xi^{-1}M^2)^2}
 =
       \frac{1}{p^2- \xi^{-1}M^2} 
        + \frac{(\xi^{-1}-\eta^{-1})M^2}{{(p^2-\xi^{-1}M^2)}^2}.
\notag
\\
 \langle \chi B \rangle =& \,
         - \frac{M}
             {p^2 - \xi^{-1}M^2},
\notag
\\
 \langle B B \rangle =&0.
 \label{propagators:Yao}
\end{align}
which shows the characteristic features of the R$_\xi$ gauge:
\begin{enumerate}
 \item for finite $\xi$, 
       the unphysical Goldstone mode $\chi$ propagates with 
        finite mass and
        the ultraviolet behavior of 
        the propagator $\langle A_\mu A_\nu \rangle$ is $O(1/p^2)$. 
 \item in the limit of $\xi \to 0$, the theory becomes that of the unitary gauge: 
        the mass of the Goldstone mode becomes infinitely large and does not 
        propagate any more, and the second term of 
        $\langle A_\mu A_\nu\rangle$
        vanishes so that 
        $\langle A_\mu A_\nu \rangle$ becomes the propagator of 
        the Proca field (see Eq.(\ref{propagator:UU})) 
        whose ultraviolet behavior is $O(1)$.
\end{enumerate}

\section{General gauge fixing  including gaugeon fields}
Now we consider the general gauge-fixing Lagrangian 
$\mathcal L_\mathrm{GF}$
which 
includes gaugeon fields $Y$ and $Y_*$. 
To this end, we use a polar decomposition of $\varphi$ field
rather than (\ref{phi:Re+Im}). 
If we use the parameterization (\ref{phi:Re+Im}) to construct a 
gaugeon formalism for the R$_\xi$ gauge, it is inevitable 
to introduce non-polynomial terms in the Lagrangian. 
To avoid this, we use the following parameterization for $\varphi$:
\begin{equation}
  \varphi (x)=\frac{1}{\sqrt 2} 
                \left( v+\rho (x) \right)
                   e^{ i \pi(x) /v },
 \label{phi:plolar}
\end{equation}
where hermitian fields $\rho(x)$ and $\pi(x)$ correspond 
to the fields $\psi(x)$ and $\chi(x)$ of (\ref{phi:Re+Im}), 
respectively; $\pi(x)$ is the Goldstone mode. 
%
In terms of these polar variables, $\mathcal{L}_\mathrm{cl}$ 
is expressed by 
\begin{align}
\mathcal{L}_\mathrm{cl} 
 = & 
     - \frac{1}{4} F_{\mu \nu} F^{\mu \nu} 
     + \frac{1}{2} M^2 \left( 1 + \frac{e}{M} \,\rho \right)^2 
                       \left( A_\mu - \frac{1}{M}\partial_\mu \pi \right)^2
   \notag \\  &
     + \frac{1}{2} \left(
                          \partial_\mu \rho \, \partial^\mu \rho 
                          - m^2 \rho^2 
                   \right) 
     - \frac{1}{2} m\sqrt{\lambda} \rho^3 
     - \frac{\lambda}{8} \rho^4
     + \frac{1}{8} m^2 v^2, 
 \label{L:cl}
\end{align}
where 
$M=ev$ is the mass of $A_\mu$ and 
$m=\sqrt{\lambda v^2}$ is the mass of the Higgs boson $\rho$.
The gauge transformation (\ref{transf:gauge}) is now 
\begin{align}
   A_\mu(x) & \to A_\mu(x) + \partial_\mu \Lambda(x), 
   \notag \\
   \pi (x)   & \to \pi (x) + M\Lambda(x), 
   \notag \\
   \rho(x)   & \to \rho (x), 
\end{align}
under which the Lagrangian 
(\ref{L:cl}) 
is invariant. 

\subsection{general gauge-fixing Lagrangian}
We impose the following conditions on the gauge-fixing Lagrangian
$\mathcal L_\mathrm{GF}$:
\begin{enumerate}
 \def\theenumi{\alph{enumi}} \def\labelenumi{(\theenumi)}  %
 \item 
   Lorentz invariance.
 \item
   quadratic in the fields $A_\mu$, $\pi$, 
   $B$, $Y$ and $Y_*$. 
 \item 
   The mass dimension\footnote{%
                Note that the mass dimensions of $B$, $Y_*$ and $Y$ 
                are different from those of usual fields; 
                $B$ and $Y_*$ have dimension two 
                while  $Y$ has dimension zero.
                               }
    of each term does not exceed four. 
    For example, we do not include such terms as 
    $\partial_\mu Y_* \partial^\mu \pi$, which has dimension five. 
    Dimension six operators such as $\partial_\mu B \partial^\mu Y_*$ 
    are also excluded from $\mathcal L_\mathrm{GF}$. 
  \item 
    BRST invariance (by incorporating  suitable FP ghost terms).
    This condition excludes those terms 
    expressed as a product 
    of two BRST parent fields, such as 
    $\partial_\mu Y \partial^\mu \pi$. 
\end{enumerate}
The most general gauge-fixing Lagrangian satisfying these conditions 
can be written by
\begin{align}
\mathcal{L}_\mathrm{GF} 
     = &
        -(\omega_1 \partial _\mu B +\omega_3 \partial_\mu Y_*) A^\mu 
        -(\omega_2 \partial_\mu B +\omega_4 \partial_\mu Y_*) \partial^\mu Y
        +(\beta_1 B + \beta_3 Y_*) M \pi 
       \notag \\
       &
        +(\beta_2 B +\beta_4 Y_*) M^2 Y
        + \frac{1}{2} \alpha_1 B^2
        + \alpha_2 B Y_* 
        + \frac{1}{2} \alpha_3 Y_*^2, 
   \label{L:general} 
\end{align}
where $\omega_i$, $\beta_i$ and $\alpha_i$ are numerical 
parameters. In order to ensure that the $\mathcal L_\mathrm{GF}$ 
properly fixes the gauge, these  parameters should satisfy at least one 
of the following three conditions:
\begin{align}
  &\omega_1 \omega_4 -\omega_2 \omega_3 \neq 0, 
  \notag \\
  &\omega_1 \beta_4 +\beta_1 \omega_4 - \omega_2 \beta_3 - \beta_2 \omega_3 
  \neq 0, 
  \notag \\
  &
  \beta_1 \beta_4 -\beta_2 \beta_3 \neq 0.  
 \label{conditions4w-b} 
\end{align}
If all these three condition are not satisfied,  we cannot obtain 
the propagators from the Lagrangian 
$\mathcal L_\mathrm{cl}+\mathcal L_\mathrm{GF}$. 

If we put $\omega_1=\omega_3=0$ and thus the first term of 
(\ref{L:general}) vanishes, then the gauge-fixing Lagrangian 
$\mathcal L_\mathrm{GF}$ 
leads to the unitary gauge, and thus 
the ultraviolet behavior of the propagators 
$\langle A_\mu A_\nu \rangle$ 
becomes $O(1)$ 
rather than $O(1/p^2)$. 
To avoid this case, 
we assume that 
at least one of the parameters $\omega_1$ and $\omega_3$ is 
not equal to zero, or equivalently, 
${\omega_1}^2+{\omega_3}^2 \neq 0$. 

Now we can consider the field redefinition of $B$ and $Y_*$
\begin{equation}
  \begin{pmatrix}
       B^\prime \\
       Y_*^\prime 
  \end{pmatrix}
  =
  \begin{pmatrix}
      \omega_1 & \omega_3 \\
      \omega_3 & -\omega_1
  \end{pmatrix}
  \begin{pmatrix}
      B \\
      Y_*
  \end{pmatrix},  
\label{B'=wB}
\end{equation}
by which the first term of (\ref{L:general}) is transformed to 
$-B^\prime \partial_\mu A^\mu$; 
the matrix in (\ref{B'=wB}) is invertible 
because of the assumption ${\omega_1}^2+{\omega_3}^2 \neq 0$. 
Thus, 
without loss of generality, 
we can assume $\omega_1=1$ and $\omega_3=0$ in (\ref{L:general}) 
owing to this field redefinition. 

Next, we consider the following field redefinitions:
\begin{align}
        A_\mu^\prime &=A_\mu +\omega_2 \partial_\mu Y, \notag \\
        \pi^\prime &=\pi +\omega_2 MY, 
\end{align}
by which the second term of 
(\ref{L:general}) 
is transformed to 
$
-\omega_4 \partial_\mu Y_* \partial^\mu Y 
$, 
while $\mathcal L_\mathrm{cl}$ 
remains invariant. Thus, without loss of generality, 
we can put $\omega_2=0$ in  (\ref{L:general}). 

Here, we further assume that $\omega_4 =1$ from the following reason. 
If $\omega_4=0$ (in addition to $\omega_2=0$),  
the second term of 
(\ref{L:general}) 
vanishes. 
Then, 
the field equations for the fields $Y$ and $Y_*$ become algebraic ones, 
and thus these fields 
should be eliminated from the Lagrangian. Namely, 
$\mathcal L_\mathrm{GF}$ no longer admits 
$q$-number gauge transformations. 
To exclude this situation, we assume $\omega_4 \neq 0$. 
With this assumption, $Y$ can be rescaled as   $\omega_4 Y \to Y$ 
to absorb the parameter $\omega_4$. In terms of the rescaled 
field, the value of $\omega_4$ is  equal to one.

Our general gauge-fixing  Lagrangian 
is now given  by 
\begin{align}
\mathcal{L}_\mathrm{GF} 
     = &
        -\partial _\mu B  A^\mu 
        -\partial_\mu Y_* \partial^\mu Y
        +(\beta_1 B + \beta_3 Y_*) M \pi 
       \notag \\ &
        +(\beta_2 B +\beta_4 Y_*) M^2 Y
        + \frac{1}{2} \alpha_1 B^2
        + \alpha_2 B Y_* 
        + \frac{1}{2} \alpha_3 Y_*^2. 
   \label{L:general2} 
\end{align}
In the following,
we also use a matrix notation to express 
this Lagrangian 
as 
\begin{equation}
\mathcal{L}_\mathrm{GF} 
     = 
        -\partial _\mu \mathcal B^\mathrm{T}  \mathcal{A}^\mu 
        + {\mathcal B}^\mathrm{T} M {{\beta}}\, \Pi 
        + \frac{1}{2}\, \mathcal B^\mathrm{T} {{\alpha}} \,\mathcal B,
   \label{L:general2'} 
\end{equation}
where 
$\mathcal A_\mu$, 
$\mathcal B$, and $\Pi$ 
denote column matrices 
defined by 
\begin{equation}
  \mathcal A_\mu = \begin{pmatrix}
                      A_\mu \\
                      \partial_\mu Y
                   \end{pmatrix}, 
 \qquad                   
      \mathcal B = \begin{pmatrix}
                      B \\
                      Y_*
                   \end{pmatrix}, 
 \qquad                   
              \Pi = \begin{pmatrix}
                      \pi \\
                      MY
                   \end{pmatrix}, 
  \label{matrix_notation1}
\end{equation}
$ \alpha$ and $ \beta$ are 
$2\times 2$ matrices defined by 
\begin{equation}
      { \alpha} = \begin{pmatrix}
                                \alpha_1 & \alpha_2 \\
                                \alpha_2 & \alpha_3
                             \end{pmatrix}, 
 \qquad                   
      { \beta} = \begin{pmatrix}
                                \beta_1 & \beta_2 \\
                                \beta_3 & \beta_4 
                            \end{pmatrix}, 
  \label{matrix_notation2}
\end{equation}
and T represents the matrix transpose. 

Note here that the Lagrangian 
(\ref{L:general2}) or (\ref{L:general2'})
satisfies the first condition 
of (\ref{conditions4w-b}) and that no more constraints on 
$\alpha_i$ and $\beta_i$ are necessary.  
Consequently, we may consider
the case such that all the $\beta_i$  
are equal to zero ($ \beta =0$); 
in this case, the gauge-fixing  Lagrangian 
is identical with that of Lorenz gauge 
by Yokoyama and Kubo (\ref{L:YK}) 
(with $\alpha_i$ chosen appropriately). 
We can also take another case that $\alpha_2=\beta_2=\beta_3=0$ 
(both $ \alpha$ and $ \beta$ are diagonal); 
in this case, the gaugeon sector of $\mathcal L_\mathrm{GF}$ decouples 
and the rest of the Lagrangian 
become identical 
to that of Yao's R$_\xi$ gauge (\ref{L:Yao}).

\section{BRST symmetric theory}
We introduce FP ghost term 
$\mathcal{L}_\mathrm{FP}$ 
into the general gauge-fixing Lagrangian 
$\mathcal{L}_\mathrm{GF}$ 
given by 
(\ref{L:general2}) or (\ref{L:general2'}).
\begin{align}
  \mathcal{L}_\mathrm{GF+FP} 
        =& \mathcal{L}_\mathrm{GF} + \mathcal{L}_\mathrm{FP} 
        \notag \\
        =& 
           - \partial_\mu  B A^\mu  - \partial_\mu Y_* \partial^\mu Y
           + (\beta_1 B +\beta_3 Y_*) M \pi 
           + (\beta_2 B + \beta_4 Y_* )M^2 Y 
       \notag \\ &
           +\frac{1}{2} \alpha_1 B^2 
           +\alpha_2 B Y_*  
           + \frac{1}{2} \alpha_3 Y_*^2 
       \notag \\
          & -i \partial _\mu {c_*} \partial ^\mu c 
             -i \partial _\mu {K_*} \partial ^\mu K
             + i (\beta_1 c_* + \beta_3 K_*)M^2 c
            + i (\beta_2 c_* + \beta_4 K_* )M^2 K, 
  \label{L:gf+fp}
\end{align}
where $c$ and $c_*$ are usual FP ghost fields and $K$ and $K_*$ are 
the FP ghost fields for gaugeon fields. 
In the matrix notation, the Lagrangian can be expressed as
\begin{align}
\mathcal{L}_\mathrm{GF+FP} 
     = &  
        -\partial _\mu \mathcal B^\mathrm{T}  \mathcal{A}^\mu 
        + {\mathcal B}^\mathrm{T} M {{\beta}}\, \Pi 
        + \frac{1}{2}\, \mathcal B^\mathrm{T} {{\alpha}} \,\mathcal B
     \notag \\ &
        - i \partial_\mu \mathcal C_*^\mathrm{T} \partial^\mu \mathcal C
        + i \mathcal C_*^\mathrm{T} M^2{{\beta}}\, \mathcal C, 
   \label{L:gf+fp'} 
\end{align}
where 
$\mathcal C$ and  
$\mathcal C_*$
denote column matrices 
defined by 
\begin{equation}
   \mathcal C = \begin{pmatrix}
                      c \\
                      K
                   \end{pmatrix}, 
 \qquad                   
   \mathcal C_*  = \begin{pmatrix}
                      c_* \\
                      K_*
                   \end{pmatrix}. 
  \label{matrix_notation_fp}
\end{equation}

\subsection{field equations}

Field equations derived from the Lagrangian 
$\mathcal L_\mathrm{cl} + \mathcal L_\mathrm{GF+FP}$ 
are, for non-FP-ghost fields,  
\begin{align}
  &
    \partial_\mu F^{\mu \nu} 
    + M^2 (1+\frac{e}{M} \rho )^2 (A^\nu -\frac{1}{M} \partial^\nu \pi )
    - \partial^\nu B 
    =0, 
 \\
  &
    \partial_\mu \left\{ 
                        M\left( 1+\frac{e}{M} \rho \right)^2 
                        \left( A^\mu -\frac{1}{M} \partial^\mu \pi \right)
                 \right\} 
    + \beta_1 M B
    + \beta_3 MY_* 
    = 0, 
  \label{eq:divProca}
  \\
   &
     (\square +m^2 )\rho + \frac{3}{2}\,m\sqrt{\lambda} \rho^2 + \frac{\lambda}{2} \rho^3 
     - 
       eM \left( 1+\frac{e}{M} \rho \right)
          \left( A_\mu -\frac{1}{M} \partial_\mu \pi \right)^2 
    = 0, 
  \\
   &
    \partial_\mu A^\mu 
    + \beta_1 M\pi + \alpha_2 Y_* 
    + \alpha_1 B   + \beta_2 M^2 Y
    = 0, 
   \label{eq:delA}
  \\
   &
    (\square + \beta_4 M^2 )Y
    + \alpha_3 Y_*  +  \alpha_2 B  + \beta_3 M \pi 
    = 0, 
   \label{eq:BoxY}
  \\
   &
    (\square + \beta_4 M^2 )Y_* + \beta_2 M^2 B 
    = 0,
   \label{eq:BoxY*}
\end{align}
from which we also have
\begin{equation}
   (\square + \beta_1 M^2 ) B + \beta_3 M^2 Y_* =0.
  \label{eq:BoxB}
\end{equation}
In the matrix notation 
(\ref{matrix_notation1}) and (\ref{matrix_notation2}), 
the equations (\ref{eq:delA}) and (\ref{eq:BoxY}) can be written by
\begin{equation}
   \partial_\mu \mathcal A^\mu 
   + M {\beta} \Pi 
   + {\alpha} \mathcal B
   =0,
  \label{eq:del_calA}
\end{equation}
and the equations 
(\ref{eq:BoxY*}) and (\ref{eq:BoxB}) by
\begin{equation}
   ( \square + M^2 {\beta}^\mathrm{T}  ) \mathcal B =0. 
  \label{eq:Box_calB}
\end{equation}
Field equations for FP ghost fields are given by
\begin{align}
  &
    (\square + M^2 {{\beta}}) \,\mathcal C=0 ,
  \label{eq:Box_calC}
  \\
  &
   ( \square + M^2 {\beta}^\mathrm{T}  ) \,\mathcal C_* =0, 
\end{align}
where we have used the matrix notation 
(\ref{matrix_notation_fp}) and (\ref{matrix_notation2}). 

The Proca field $U_\mu$ can be defined by
\begin{equation}
   U_\mu = A_\mu - \frac{1}{M}\partial_\mu \pi 
                 - \frac{1}{M^2}\partial_\mu B,
  \label{def:Proca}
\end{equation}
which satisfies
\begin{align}
     &
     (\square + M^2) U_\mu=0, 
     \\
     &
     \partial_\mu U^\mu =0, 
\end{align}
where we have assumed the free field approximation, that is, 
we consider the case $e \to 0$ 
but $M \neq 0$.

In the free field approximation, 
(\ref{eq:divProca}) and (\ref{eq:delA}) lead to the field equation for
the $\pi$ field: 
\begin{equation}
  (\square + M^2 \beta_1 ) \pi + M^3 \beta_2 Y
     +M(\alpha_1 - \beta_1)B + M(\alpha_2 - \beta_3)Y_*=0, 
  \label{eq:Boxpi}
\end{equation}
which, 
together with (\ref{eq:BoxY}), may be 
expressed as
\begin{equation}
   ( \square + M^2 {\beta} ) \Pi 
     + M (\alpha - E_{(11)} \beta^\mathrm{T}) \mathcal B =0, 
  \label{eq:BoxPi}
\end{equation}
where $E_{(11)}$ is  a matrix defined by 
\begin{equation} 
   E_{(11)} = \begin{pmatrix}
           1 & 0\\
           0 & 0
         \end{pmatrix}.
\label{def:E_1}         
\end{equation}

\subsection{BRST symmetry}

Our Lagrangian 
$\mathcal L_\mathrm{cl} + \mathcal L_\mathrm{GF+FP}$ 
is invariant under the BRST transformation,  
\begin{align}
    \delta_\mathrm{B} A_\mu &= \partial_\mu c, 
    \qquad \delta_\mathrm{B} \pi = Mc,
    \qquad \delta_\mathrm{B} \rho =0,
    \notag \\
    \delta_\mathrm{B} {c_*} &= iB, 
    \notag \\
    \delta_\mathrm{B} B &= \delta_\mathrm{B} c =0,
    \notag \\
    \delta_\mathrm{B}Y &= K,
    \notag \\
    \delta_\mathrm{B} K_* &= iY_*,
    \notag \\
    \delta_\mathrm{B} Y_* &= \delta_\mathrm{B} K=0, 
   \label{delta_B1} 
\end{align}
which can be also expressed as
\begin{align}
    \delta_\mathrm{B} \mathcal A_\mu & = \partial_\mu \mathcal C, 
    \qquad 
    \delta_\mathrm{B} \Pi = M \mathcal C,
    \qquad \delta_\mathrm{B} \rho =0,
    \notag \\
    \delta_\mathrm{B} \mathcal C_* &= i\mathcal B, 
    \notag \\
    \delta_\mathrm{B} \mathcal B &= \delta_\mathrm{B} \mathcal C =0.
   \label{delta_B1'} 
\end{align}
This obviously satisfies the nilpotency, ${\delta_\mathrm{B}}^2=0$. 
Because of the nilpotency, the BRST invariance of $\mathcal L_\mathrm{GF+FP}$ 
can be easily seen if we rewrite the Lagrangian as
\begin{equation}
    \mathcal L_\mathrm{GF+FP}
    =
    -i \delta_\mathrm{B} 
          \left[ 
               - \partial _\mu \mathcal C_*^\mathrm{T} \mathcal A^\mu
               + \mathcal C_*^\mathrm{T}
                           \left( M {\beta} \Pi 
                                  + \frac{1}{2} {\alpha} \mathcal B
                          \right)
         \right].                
\end{equation}
The corresponding BRST current $J_\mathrm{B}^\mu$ is given by
\begin{align}
  J_\mathrm{B}^\mu  &= B \overleftrightarrow{\partial ^\mu} c 
                    + Y_* \overleftrightarrow{\partial ^\mu} K
  \notag \\ &
                    = \mathcal B^\mathrm{T} \overleftrightarrow{\partial ^\mu} \mathcal C. 
\end{align}
Conservation of this current can be easily seen  as 
\begin{equation}
  \partial_\mu J_\mathrm{B}^\mu  
           = 
           \mathcal B^\mathrm{T} (\overrightarrow{\square} 
                         -  \overleftarrow{\square} )\mathcal C
          =0, 
\end{equation}
where we have used the 
field equations (\ref{eq:Box_calB}) and (\ref{eq:Box_calC}).

The corresponding BRST charge is thus given by
\begin{align}
Q_\mathrm{B} =
\int d^3 x  \mathcal B^\mathrm{T} \overleftrightarrow{\partial ^0} \mathcal C
=
\int d^3 x[B\overleftrightarrow{\partial ^0} c
+Y_* \overleftrightarrow{\partial ^0} K ]. 
\label{Q_B}
\end{align}
By the help of this charge we can define the physical subspace 
$\mathcal V_\mathrm{phys}$ as a space of 
states satisfying 
\begin{align}
                 Q_\mathrm{B} |\mathrm{phys} \rangle =0. 
     \label{Q_Bphys}
\end{align}
By using this subsidiary condition, we can remove 
all unphysical modes of this theory.\cite{K-O,Kugo} 

Note that the Proca field $U_\mu$ defined by (\ref{def:Proca}) is 
BRST invariant: 
\begin{equation}
          \delta_\mathrm{B} U_\mu = 0.
\end{equation}

\subsection{$q$-number gauge transformation}

We define  $q$-number gauge transformations  by
\begin{align}
    A_\mu &\to \hat{A}_\mu =A_\mu +\tau \partial_\mu Y,
    \notag 
    \\
    \pi &\to \hat{\pi} = \pi +\tau MY, 
    \notag 
    \\
    Y_* &\to \hat Y_* = Y_* -\tau B,
    \notag 
    \\
    B &\to \hat B, \qquad Y \to \hat Y, 
    \notag 
    \\
    c &\to \hat{c} =c+\tau K, 
    \notag 
    \\
    K_* &\to \hat K_* = K_* - \tau c_*, 
    \notag 
    \\
    c_* &\to \hat c_* =c_*, \qquad K \to \hat K = K, 
\label{Eq:q-transf1}
\end{align}
where $\tau$ is a numerical parameter.  
We also define the transformation for the physical Higgs field $\rho$, 
which should be invariant:
\begin{equation}
       \rho \to \hat{\rho} =\rho. 
\end{equation}
Note that the Proca filed $U_\mu$ is also 
invariant under the transformation:
\begin{equation}
    U_\mu \to \hat U_\mu = U_\mu.
\end{equation}
If we introduce a one-parameter matrix $g(\tau)$ by
\begin{equation}
   g(\tau) = \begin{pmatrix}
                            1 & \tau  
                            \\
                            0  & 1
                         \end{pmatrix},
\end{equation}
the $q$-number gauge transformed fields in (\ref{Eq:q-transf1}) 
can be expressed as
\begin{align}
   & \hat{\mathcal A}_\mu 
    =  g(\tau) \,\mathcal A_\mu ,
    \qquad
    \hat{\Pi} = g(\tau) \,\Pi,
    \\
    &\hat{\mathcal B} ^\mathrm{T} 
     = \mathcal B ^\mathrm{T} g(\tau)^{-1},
    \\
    &\hat{\mathcal C} 
    = g(\tau) \,\mathcal C,
     \qquad
    \hat{\mathcal C_*}^{T} = \mathcal {C_*} ^\mathrm{T} g(\tau)^{-1}, 
\end{align}
where $g(\tau)^{-1}=g(-\tau)$ is the inverse 
matrix of $g(\tau)$. 

Under the $q$-number gauge transformation $\mathcal L_\mathrm{cl}$ is invariant 
while 
$\mathcal{L}_{\mathrm{GF}+\mathrm{FP}}$ 
transforms as 
\begin{align}
\mathcal{L}_{\mathrm{GF}+\mathrm{FP}} 
     =&
        - \partial_\mu \hat{\mathcal B}^\mathrm{T}  \hat{\mathcal A}^\mu 
        + \hat{\mathcal B}^\mathrm{T} M \hat{{\beta}}\, \hat{\Pi} 
        + \frac{1}{2}\, \hat{\mathcal B}^\mathrm{T} \hat{{\alpha}} \,\hat{\mathcal B}
     \notag \\ &
        - i \partial_\mu \hat{\mathcal C_*}^\mathrm{T} \partial^\mu \hat{\mathcal C}
        + i \hat{\mathcal C_*}^\mathrm{T} M^2 \hat{{\beta}}\, \hat{\mathcal C}, 
   \label{L:gf+fp_roof} 
\end{align}
with $\hat{ \alpha}$ and 
$\hat{ \beta}$ 
being
\begin{align}
    \hat{{ \alpha}} 
    &= g(\tau) \,{\alpha} \,g(\tau)^\mathrm{T},
    \label{transf:alpha}
    \\
    \hat{ \beta} 
    &= g(\tau) \, \beta \,g(\tau)^{-1}.
    \label{transf:beta}
\end{align}
Thus, under the $q$-number gauge transformation, 
the total Lagrangian 
$\mathcal L= \mathcal L_\mathrm{cl}+\mathcal L_\mathrm{GF+FP}$ 
is form invariant:
\begin{equation}
  \mathcal{L} (\phi_A, \alpha_i, \beta_j )
  = \mathcal{L} (\hat{\phi}_A, \hat{\alpha}_i, \hat{\beta}_j ), 
\end{equation}
where $\hat\alpha_i$ and $\hat\beta_j$ are components of the matrices 
$\hat{ \alpha}$ and $\hat{ \beta}$, that is, 
\begin{align}
     \hat{\alpha}_1 &=\alpha_1 +2\alpha_2 \tau +\alpha_3 \tau^2, 
       \notag \\
     \hat{\alpha}_2 &=\alpha_2 +\alpha_3 \tau, \quad 
       \notag \\
     \hat{\alpha}_3 &=\alpha_3,  
       \notag \\
     \hat{\beta}_1 &=\beta_1 +\beta_3 \tau, 
       \notag \\
     \hat{\beta}_2 &=\beta_2 +(\beta_4 -\beta_1 )\tau -\beta_3 \tau^2 
       \notag \\
     \hat{\beta}_3 &=\beta_3, \quad 
       \notag \\
     \hat{\beta}_4 &=\beta_4 -\beta_3 \tau. 
\end{align}

We emphasize here that 
this $q$-number gauge transformation commutes with 
the BRST transformation (\ref{delta_B1}). 
As a result, the BRST charge (\ref{Q_B}) is invariant 
under the $q$-number gauge transformation, 
\begin{align}
        \hat{Q}_\mathrm{B} =Q_\mathrm{B},
\end{align}
and therefore the physical subspace is also invariant:
\begin{equation}
  \hat {\mathcal V}_\mathrm{phys} = \mathcal V_\mathrm{phys}.
\end{equation}


\subsection{free propagators}

The quadratic part of our Lagrangian 
$\mathcal L_\mathrm{cl} + \mathcal L_\mathrm{GF+FP}$
can be written as 
\begin{align}
  \mathcal L_\mathrm{quadratic}
  =&
  -\frac{1}{4}F_{\mu\nu}F^{\mu\nu} + \frac{M^2}{2} U_\mu U^\mu 
  + \frac{1}{2} \partial_\mu \rho \, \partial^\mu \rho - \frac{m^2}{2} \rho^2 
  \notag 
  \\
  &
  - \frac{1}{M} \partial_\mu \mathcal B^\mathrm{T} \partial^\mu \Pi
  + \mathcal B^\mathrm{T} M\beta  \Pi
  - \frac{1}{2M^2} \partial_\mu \mathcal B^\mathrm{T} E_{(11)} \partial^\mu \mathcal B
  + \frac{1}{2} \mathcal B^\mathrm{T} \alpha \, \mathcal B
  \notag
  \\
  &
  - i \partial _\mu \mathcal C_*^\mathrm{T} \partial ^\mu \mathcal C
  + i \mathcal C_*^\mathrm{T} M^2 \beta \,\mathcal C,
\end{align}
where $U_\mu$ is the Proca field (\ref{def:Proca}) 
and $E_{(11)}$ is the matrix 
defied by (\ref{def:E_1}). 
This expression shows that $U_\mu$ and $\rho$ decouple from 
other fields $\Pi$, $\mathcal B$, $\mathcal C$ and $\mathcal C_*$; 
$F_{\mu\nu}$ can be also expressed as 
$F_{\mu\nu}=\partial_\mu U_\nu - \partial_\nu U_\mu$. 
The propagators among $U_\mu$'s and $\rho$ is thus given by
\begin{align}
 &  \langle \rho \rho \rangle = \frac{1}{p^2 - m^2 } ,
 \\
 &  \langle U_\mu U_\nu \rangle 
      = \frac{1}{p^2 - M^2}
        \left( - g_{\mu\nu} + \frac{p_\mu p_\nu}{M^2}
        \right). 
    \label{propagator:UU}
\end{align}
Any other propagators which include $U_\mu$ or $\rho$ are equal to zero. 

Before discussing the propagators for other fields, we first define the 
quantity, 
\begin{align}
  D(p^2;\beta) &= \det (-p^2 \boldsymbol 1 + M^2 \beta) 
  \notag \\
               &= (p^2)^2 - p^2 M^2 t + M^4 d ,
\end{align}
where $\boldsymbol 1 $ stands for the unit matrix and 
\begin{align}
  t&= \mathrm{tr}\, \beta = \beta_1 + \beta_4,
  \\
  d&= \det \beta = \beta_1 \beta_4 - \beta_2 \beta_3,
\end{align}
Then, we get
\begin{align}
  (-p^2 \boldsymbol{1} + M^2 \beta) ^{-1} 
                    &=  \frac{1}{D(p^2;\beta)} (-p^2 \boldsymbol{1} + M^2 \tilde \beta ), 
\end{align}
where $\tilde \beta$ is the cofactor matrix of $\beta$, 
\begin{equation}
  \tilde \beta = \begin{pmatrix}
                      \beta_4 & -\beta_2
                      \\
                      - \beta_3 & \beta_1
                 \end{pmatrix},
\end{equation}
which satisfies 
\begin{align}
   &\beta \tilde \beta = \tilde \beta \beta = \det \beta \cdot \boldsymbol{1},
   \\
   & \beta + \tilde \beta = \mathrm{tr}\, \beta \cdot \boldsymbol{1}.
\end{align}
Note that $\tilde \beta $ transforms 
as 
$ \hat {\tilde \beta} \equiv \tilde{\hat \beta} = g(\tau) \tilde \beta g(\tau)^{-1}$ 
under the $q$-number gauge transformation. 

The propagators among $\Pi$, $\mathcal B$, $\mathcal C$ and $\mathcal C_*$
are now given by 
\begin{align}
   \langle \Pi \, \Pi^\mathrm{T} \rangle 
   =&
       \frac{1}{\{D(p^2; \beta)\}^2}
         (-p^2 \boldsymbol{1} +M^2 \tilde \beta)
            (p^2 E_{(11)} -M^2 \alpha)
               (-p^2 \boldsymbol{1} +M^2 \tilde \beta)^\mathrm{T}
   \notag 
   \\
   =& 
       \frac{1}{D(p^2; \beta)}
                 (p^2 E_{(11)} - M^2 \alpha^{\prime})
       +
       \frac{M^4}{\{D(p^2; \beta)\}^2}
                 (p^2 \gamma  - M^2 \delta 
                 ),
 \label{propagator:PiPi}
 \\
  \langle \Pi \, \mathcal B^\mathrm{T} \rangle 
      =& \frac{M}{D(p^2; \beta)} (-p^2 \boldsymbol{1} + M^2 \tilde \beta),
 \label{propagator:PiB}
 \\
  \langle \mathcal B \, \mathcal B^\mathrm{T} \rangle 
     =& 0,
 \label{propagator:BB}
 \\
  \langle \mathcal C \, \mathcal C_*^\mathrm{T} \rangle 
      =& \frac{-i}{D(p^2; \beta)} (-p^2 \boldsymbol{1} + M^2 \tilde \beta),
\end{align}
where 
$\alpha^\prime$, $\gamma$ and $\delta$ 
in (\ref{propagator:PiPi}) 
are symmetric matrices 
defined by 
\begin{align}
   \alpha^\prime =& 
               \alpha + \tilde \beta E_{(11)} + E_{(11)} \tilde \beta^\mathrm{T} 
               - t E_{(11)} 
               =
                  \begin{pmatrix}
                      \alpha_1^\prime & \alpha_2^\prime
                      \\
                      \alpha_2^\prime & \alpha_3^\prime
                  \end{pmatrix},
\\
   \gamma =& 
             -t \alpha^\prime 
             -d E_{(11)} 
             + \tilde{\beta} \alpha  + \alpha \tilde{\beta}^\mathrm{T}
             + \tilde{\beta} E_{(11)}  \tilde{\beta}^\mathrm{T}
            =
               \begin{pmatrix}
                  \gamma _1 & \gamma _2
                  \\
                  \gamma _2 & \gamma _3
               \end{pmatrix},
\\
%
   \delta =& 
              - d \alpha^\prime 
                + \tilde{\beta} \alpha  \tilde{\beta}^\mathrm{T}
              =
               \begin{pmatrix}
                  \delta _1 & \delta _2
                  \\
                  \delta _2 & \delta _3
               \end{pmatrix},
\end{align}
which lead to 
\begin{align}
 &
 \left\{
  \begin{array}{l}
   \alpha_1^\prime = \alpha_1 - \beta_1 + \beta _4, 
   \\
   \alpha_2^\prime = \alpha_2 - \beta_3, 
   \\
   \alpha_3^\prime = \alpha_3,
  \end{array}
 \right.     
 \\
 &
 \left\{
  \begin{array}{l}
   \gamma_1 = -\alpha_1(\beta_1-\beta_4) - 2\alpha_2 \beta_2 + {\beta_1}^2 -d ,
   \\
   \gamma_2 = -\alpha_1 \beta_3 -\alpha_3 \beta_2 + \beta_1 \beta_3,
   \\
   \gamma_3 = -2 \alpha_2 \beta_3 + \alpha_3(\beta_1 - \beta_4) + {\beta_3}^2,
  \end{array}
 \right.     
 \\
 &
 \left\{
  \begin{array}{l}
     \delta_1 = \alpha_1(-d + {\beta_4}^2) -2 \alpha_2 \beta_2 \beta_4 
                     + \alpha_3 {\beta_2}^2 
                     + (\beta_1 - \beta_4 )d, 
   \\
     \delta_2 = -\alpha_1 \beta_3 \beta_4 + 2 \alpha_2 \beta_2 \beta_3 
                      - \alpha_3 \beta_1 \beta_2
                      + \beta_3 d,
   \\
     \delta_3 = \alpha_1 {\beta_3}^2 - 2 \alpha_2 \beta_1 \beta_3
                      - \alpha_3 ( -d + {\beta_1}^2).
  \end{array}
 \right. 
\end{align}
It can be easily confirmed that under the $q$-number gauge transformation 
(\ref{transf:alpha}) and
(\ref{transf:beta})
the matrices $\alpha^\prime$, $\gamma$ and $\delta$ 
transform as 
$\hat{\alpha}^\prime =g(\tau)\alpha^\prime g(\tau)^\mathrm{T}$, 
$\hat{\gamma} =g(\tau)\gamma g(\tau)^\mathrm{T}$ and
$\hat{\delta} =g(\tau)\,\delta\, g(\tau)^\mathrm{T}$.

Since $A_\mu=U_\mu+\partial_\mu \pi/M+ \partial_\mu B/M^2$, the propagators 
for $A_\mu$'s can be evaluated from the propagators for other fields, 
such as, 
\begin{align}
  \langle A_\mu A_\nu \rangle 
     =& 
        \langle U_\mu U_\nu \rangle 
      + \frac{p_\mu p_\nu }{M^2}
           \left[
               \langle \pi \pi \rangle 
             + \frac{1}{M} \big(\langle \pi B \rangle + \langle B \pi \rangle \big)
             + \frac{1}{M^2}  \langle BB \rangle 
           \right],
      \label{propagator:AAbyUpiB}
  \\
  \langle A_\mu \pi \rangle 
     =& 
       \frac{-ip_\mu }{M}
           \left[
               \langle \pi \pi \rangle 
             + \frac{1}{M} \langle  B \pi \rangle 
           \right].  
      \label{propagator:ApibyUpiB}
\end{align}
Thus, from (\ref{propagator:UU}) and (\ref{propagator:PiPi})-%
(\ref{propagator:BB}),  we get
\begin{align}
 \langle A_\mu A_\nu \rangle 
      =&  \frac{1}{p^2 - M^2} \left[ - g_{\mu\nu} + \frac{p_\mu p_\nu}{M^2} \right]
      \notag \\
       & + \frac{p_\mu p_\nu}{M^2}
           \left[ 
                  \frac{-p^2 + (t - \alpha_1  )M^2}{D(p^2;\beta)} 
        +M^4
                 \frac{\gamma_1 p^2 - \delta_1 M^2}{{D(p^2;\beta)}^2}
          \right],
      \label{propagator:AA1}
      \\
 \langle A_\mu \rho \rangle =&\, 0,
 \\
 \langle A_\mu \pi \rangle 
       =&\,
        i p_\mu M 
          \left[
                \frac{\alpha_1- \beta_1}{{D(p^2;\beta )}}
                -
                \frac{M^2 (\gamma_1 p^2 - \delta_1 M^2)}{{D(p^2;\beta)}^2}
          \right],
        \\
  \langle A_\mu B \rangle 
       =&\, ip_\mu \frac{p^2- \beta_4 M^2}{D(p^2;\beta)}, 
        \\
  \langle A_\mu Y \rangle 
       =&
       \, ip_\mu \left[ 
                       \frac{\alpha_2 }{D(p^2; \beta)}
                       -
                       \frac{ M^2(\gamma_2 p^2 - \delta_2 M^2)}{{D(p^2; \beta)}^2}
                 \right],
        \\
    \langle A_\mu Y_* \rangle 
       =&\,
       ip_\mu  \frac{\beta_2 M^2}{D(p^2;\beta)}.
    \label{prpagators} 
\end{align}
Note here the propagator (\ref{propagator:AA1}) reads 
\begin{align}
 \langle A_\mu A_\nu \rangle 
      =& - \frac{g_{\mu \nu}}{p^2 - M^2} 
        - (\alpha_1 -1) \frac{ p_\mu p_\nu}{D(p^2;\beta)}
        \notag \\
        &
        + M^2 p_\mu p_\nu \left[
                            \frac{1-t+d}{(p^2-M^2)D(p^2;\beta)}
                            +
                            \frac{\gamma_1 p^2 - \delta_1 M^2}{{D(p^2;\beta)}^2}
                          \right],
      \label{propagator:AA2}
\end{align}
which shows that 
$\langle A_\mu A_\nu \rangle$ 
has the usual ultraviolet behavior, 
\begin{equation}
     \langle A_\mu A_\nu \rangle = O\left(\frac{1}{p^2}\right), 
     \quad \text{as $p^\mu \to \infty$}
\end{equation}
rather than that of the Proca filed,  
$\langle U_\mu U_\nu \rangle = O(1)$. 

\section{R$_\xi$-like gauges}
Our theory includes R$_\xi$-like gauges.
By choosing the gauge-fixing parameters $\alpha$ and $\beta$ appropriately, 
we find that some of the parameters can be considered as a $\xi$ parameter.
That is, some of the parameter has a property similar to that of the $\xi$ 
parameters of R$_\xi$ gauge\cite{Fujikawa,Yao} mentioned in \S 2. 

\subsection{Yao's gauge}
We choose the parameters $\alpha$ and $\beta$ as
\begin{align}
  \alpha &= \begin{pmatrix}
             \eta^{-1} & 1/2 \\
             1/2 & 0
            \end{pmatrix},
  \qquad
  \beta = \xi^{-1} \boldsymbol{1} 
       =
           \begin{pmatrix}
             \xi^{-1} & 0\\
             0 & \xi^{-1}
            \end{pmatrix}, 
 \label{parameter_choice:Yao}
\end{align}
then 
the 
gauge-fixing term of the Lagrangian  is now
\begin{equation}
\mathcal{L}_\mathrm{GF} 
  = B\left( \partial_\mu A^\mu + \frac{1}{\xi}M \pi \right)
    + \frac{1}{2\eta} B^2 + \frac{1}{2}BY_* 
    -\partial_\mu Y_* \partial^\mu Y + \frac{1}{\xi}M^2 Y_* Y, 
\end{equation}
up to total derivatives. 
The first two terms of the right hand side are
the same gauge-fixing term as Yao's 
(\ref{L:Yao}). Furthermore, we can easily see that
$A_\mu$, $\pi\ (\approx \chi)$ and $B$ have the same propagators 
as Yao's (\ref{propagators:Yao}).

Under the $q$-number gauge transformation (\ref{Eq:q-transf1}), 
the parameters $\xi$ and $\eta$ transform  as
\begin{align}
  \xi^{-1} &\to \hat \xi^{-1} = \xi^{-1} 
  \\
  \eta^{-1} &\to \hat \eta^{-1}=\eta^{-1}+ \tau. 
\end{align}
Thus the parameter $\eta$ can be shifted freely. 
In particular, by using the 
$q$-number gauge transformation, we can put $\hat \xi= \hat \eta$; the 
theory become equivalent to that of Fujikawa-Sanda-Lee.\cite{Fujikawa}
In contrast to $\eta$, the parameter $\xi$ cannot be shifted. 
Consequently, 
we cannot take the limit $\xi \to 0$ by the $q$-number 
gauge transformation.

In addition to the  BRST charge  (\ref{Q_B}), we can define the 
following BRST-like conserved 
charges: 
\begin{align}
    & Q_\mathrm{B(KO)} = \int c \overleftrightarrow{\partial_0} B d^3x, 
\label{Q_BKO}
\\
     & Q_\mathrm{B(Y)} = \int K \overleftrightarrow{\partial_0} Y_* d^3x, 
\label{Q_BY}
\\
     & Q_\mathrm{B(KO)}^\prime = \int K \overleftrightarrow{\partial_0} B d^3x, 
\label{Q_BKO'}
\\
     & Q_\mathrm{B(Y)}^\prime = \int c \overleftrightarrow{\partial_0} Y_* d^3x. 
\label{Q_BY'}
\end{align}
All of these satisfy the nilpotency condition. 
The conservation of these charges are due to the fact that 
in the case of $\beta= \xi^{-1} \boldsymbol{1}$ 
the fields $B$, $Y_*$, $c$ 
and $K$ satisfy the same Klein-Gordon equation with the same 
mass squared $M^2/\xi$. 

Instead of the physical condition (\ref{Q_Bphys}), 
we may consider the condition 
\begin{align}
            &     Q_\mathrm{B(KO)} |\mathrm{phys} \rangle =0, 
    \notag \\
            &     Q_\mathrm{B(Y)} |\mathrm{phys} \rangle =0, 
     \label{Q_Bphys2}
\end{align}
to define the physical subspace. The unphysical Goldstone mode 
is removed by the first equation while the gaugeon modes are 
removed by the second.
 Let 
$\mathcal V_\mathrm{phys}^{(\eta)}$ denote the space of states 
satisfying (\ref{Q_Bphys2}). This space is a subspace of 
$\mathcal V_\mathrm{phys}$ defined by (\ref{Q_Bphys}): 
$
     \mathcal V_\mathrm{phys}^{(\eta)} 
     \subset 
     \mathcal V_\mathrm{phys}
$. 
The definition of the space 
$\mathcal V_\mathrm{phys}^{(\eta)}$ depends on the parameter $\eta$. 
In fact, under the $q$-number gauge transformation, the charges 
$Q_\mathrm{B(KO)}$ and 
$Q_\mathrm{B(Y)}$ transform as 
\begin{align}
   & Q_\mathrm{B(KO)} \to \hat Q_\mathrm{B(KO)} 
        =  Q_\mathrm{B(KO)} + \tau  Q_\mathrm{B(Y)}^\prime,
   \notag \\
   & Q_\mathrm{B(Y)} \to \hat Q_\mathrm{B(Y)} 
        =  Q_\mathrm{B(Y)} - \tau  Q_\mathrm{B(KO)}^\prime.
\end{align}
Consequently, 
the  subspace $\mathcal V_\mathrm{phys}^{(\eta)}$ transforms 
into another subspace 
$\mathcal V_\mathrm{phys}^{(\eta + \tau)}$. 

Let us define a subspace $\mathcal V^{(\eta)}_\mathrm{Yao}$ of the 
total Fock space $\mathcal V$ by
\begin{equation}
     \mathcal V^{(\eta)}_\mathrm{Yao} = \ker Q_\mathrm{B(Y)}
      = \{ |\Phi \rangle \in \mathcal V ; Q_\mathrm{B(Y)}|\Phi \rangle = 0\}
      \subset \mathcal V,
\end{equation}
which includes 
$\mathcal V_\mathrm{phys}^{(\eta)}$
as a subspace. Since the gaugeon modes are excluded from 
$\mathcal V^{(\eta)}_\mathrm{Yao}$, the space 
$\mathcal V^{(\eta)}_\mathrm{Yao}$
corresponds to the total Fock space of the Yao's theory. 
Under the $q$-number gauge transformation this subspace 
transforms as 
\begin{equation}
    \mathcal V^{(\eta)}_\mathrm{Yao}
    \to 
    \mathcal V^{(\eta+\tau)}_\mathrm{Yao}= \ker \hat Q_\mathrm{B(Y)}
    \subset \mathcal V.
  \label{gauge_structure}
\end{equation}
Thus various Fock spaces of the Yao's theory (including the theory of 
Fujikawa, Sanda and Lee) corresponding to various values of $\eta$ are 
embedded in the single Fock space $\mathcal V$ of our theory.%
\footnote{%
    If we put $\beta=0$ ($\xi \to \infty$), the theory becomes the 
    BRST symmetric version of the gaugeon formalism for the  Lorenz gauge 
    of Yokoyama and Kubo (\ref{L:YK}). 
    The same gauge structure of the Fock spaces as (\ref{gauge_structure})
     still holds for this theory. 
    }

\subsection{more complicated cases}
We choose the parameters $\alpha$ and $\beta$ as
\begin{align}
  \alpha &= \begin{pmatrix}
             \eta^{-1} & 1/2 \\
             1/2 & 0
            \end{pmatrix},
  \qquad
  \beta 
       =
           \begin{pmatrix}
             \xi^{-1} & \beta_2\\
             0 & \xi^{-1}
            \end{pmatrix}, 
\end{align}
then 
the determinant $D(p^2;\beta)$ and matrices
$\alpha^\prime$, $\gamma$ and $\delta$ become 
\begin{align}
  & D(p^2;\beta) = 
   (p^2-\xi^{-1}M^2)^2,
   \\
  & \alpha^\prime = \alpha, 
    \quad \gamma=-\beta_2 E_{(11)}, 
   \quad \delta= -\xi^{-1}\beta_2 E_{(11)}.
\end{align}
Free propagators for $A_\mu$ and $\pi$  
are given by 
\begin{align}
 \langle A_\mu A_\nu \rangle 
       =& \,
           \frac{1}{p^2 - M^2} \left(
                                      - g_{\mu\nu} + \frac{p_\mu p_\nu}{M^2}
                                \right)
         \notag \\ &
         +\frac{p_\mu p_\nu}{M^2}
                \left[
                     -  \frac{1}{p^2 -\xi^{-1}M^2}
                     + \frac{(\xi^{-1}-\eta^{-1})M^2}{{(p^2-\xi^{-1}M^2)}^2}
                     - \frac{\beta_2 M^4}{{(p^2-\xi^{-1}M^2)}^3}
                \right],
 \notag
\\
 \langle A_\mu \pi \rangle 
       =&\,
        -ip_\mu \left[
               \frac{(\xi^{-1}-\eta^{-1})M}{(p^2 - \xi^{-1}M^2)^2}
             - \frac{\beta_2 M^3}{(p^2 - \xi^{-1}M^2)^3}
               \right],
\notag
\\
 \langle \pi \pi \rangle 
        =& \,
       \frac{1}{p^2- \xi^{-1}M^2} 
        + \frac{(\xi^{-1}-\eta^{-1})M^2}{{(p^2-\xi^{-1}M^2)}^2}
        - \frac{\beta_2 M^4}{{(p^2-\xi^{-1}M^2)}^3}.
 \label{propagators:Ex2}
\end{align}
As easily seen, if we put $\beta_2=0$, the theory becomes Yao's gauge. 

Under the $q$-number gauge transformation, the parameter
$\xi$, $\eta$ and $\beta_2$ transforms as 
\begin{align}
  \xi^{-1} &\to \hat \xi^{-1} = \xi^{-1} ,
  \notag \\
  \eta^{-1} &\to \hat \eta^{-1}=\eta^{-1}+ \tau, 
  \notag \\
  \beta_2 &\to \hat \beta_2 = \beta_2. 
\end{align}
The $\xi$ parameter again cannot be shifted.

A bit different case is
\begin{align}
    \alpha &= \begin{pmatrix}
             \alpha_1 & 1/2 \\
             1/2 & 0
            \end{pmatrix},
  \qquad
  \beta 
       =  \begin{pmatrix}
             \beta_1 & \beta_2\\
             0 & \beta_4
            \end{pmatrix}. \qquad (\beta_1 \neq \beta_4)
\end{align}
In this case, the determinant 
$D(p^2;\beta)$ becomes 
\begin{align}
  & D(p^2;\beta) = (p^2-\beta_1 M^2)(p^2-\beta_4 M^2),
\end{align}
and it can be seen that 
${\beta_1}^{-1}$ has the same property as the  $\xi$ parameter. 
Under the $q$-number gauge transformation,  parameters $\alpha_1$,
$\beta_1$, $\beta_2$ and $\beta_4$  
transform as 
\begin{align}
    & \hat \alpha_1 = \alpha_1 + \tau, 
    \notag \\
    & \hat \beta_1 =  \beta_1
    \notag \\
    & \hat \beta_2 =  \beta_2 + (\beta_4- \beta_1) \tau, 
    \notag \\
    & \hat \beta_4 =  \beta_4.
 \label{q-transf.ex}
\end{align}
Thus, the $\xi$ parameter $\beta_1$ is again invariant. 

It should be commented that if $\beta_2=0$ the propagators among
$A_\mu$, $\pi$ and $B$ fields are the same propagators
in Yao's gauge with $\xi=\beta_1^{-1}$ and $\eta=\alpha_1^{-1}$. 
In this case, however, the $q$-number gauge transformation 
(\ref{q-transf.ex}) shifts the value of $\beta_2$ to non-zero so that 
the propagators no longer maintain the form of Yao's gauge. 

\subsection{$\xi$-parameter in general case}
We have exhibited above some examples of the R$_\xi$-like gauge which are 
included in our theory as the special choices of 
$\alpha$ and $\beta$. 
In any of these examples, the $\xi$-parameter has been  invariant under the 
$q$-number gauge transformation. 
We show here that even in more general cases the possible 
$\xi$-parameter 
is always invariant under the $q$-number gauge transformation.

First, we define the R$_\xi$ gauge of our theory as follows: 
There exits some gauge-fixing parameter(s) denoted by $\xi$ such that 
in the limit of $\xi \to 0$ the propagators among $A_\mu$'s and $\pi$ 
become the propagators of the unitary gauge, that is, 
\begin{equation}
 \left\{
 \begin{array}{l}
  \langle A_\mu A_\nu \rangle 
  \to 
  \langle U_\mu U_\nu \rangle , 
  \\
  \langle A_\mu \pi \rangle  \to 0, 
 \\
  \langle \pi \pi \rangle  \to 0, 
 \end{array}
 \right.
  \label{xi_limit:Api}
\end{equation}
where 
$  \langle U_\mu U_\nu \rangle $ 
is the propagator for the Proca field (\ref{propagator:UU}). 
Owing to the equations 
(\ref{propagator:AAbyUpiB}) and (\ref{propagator:ApibyUpiB}), 
the condition (\ref{xi_limit:Api}) can 
be read as 
\begin{equation}
 \left\{
   \begin{array}{l}
      \langle \pi \pi \rangle  \to 0, 
      \\
       \langle \pi B \rangle  \to 0,   
      \\ 
      \langle BB \rangle  \to 0.   
   \end{array}
 \right.
 \label{xi_limit:piB}
\end{equation}
We find that
from 
(\ref{propagator:PiPi})-
(\ref{propagator:BB}) 
this  condition is  equivalent to 
\begin{equation}
  D(p^2;\beta) = \det (-p^2 \boldsymbol 1 + M^2 \beta) \to \infty.
 \label{xi_limit:D(p^2)}
\end{equation}
The determinant 
$D(p^2;\beta)$ 
can be factorized as 
\begin{equation}
  D(p^2;\beta) = 
      (p^2- \xi_1^{-1} M^2)
      (p^2- \xi_2^{-1} M^2), 
\end{equation}
where 
$\xi_1^{-1}$
and 
$\xi_2^{-1}$
are two eigenvalues of the matrix $\beta$. 
The condition 
(\ref{xi_limit:D(p^2)}) shows that the possible $\xi$-parameter 
is one or both of the parameters $\xi_1$ and $\xi_2$. 
Since the eigenvalues of the matrix $\beta$ are invariant 
under the $q$-number gauge transformation (\ref{transf:beta}), 
our possible $\xi$-parameters 
$\xi_1$
and/or  
$\xi_2$
cannot be shifted by the $q$-number gauge transformation.

\section{Summary and discussion}
Starting from the most general gauge-fixing Lagrangian including $Y$ and $Y_*$ fields, 
we present a general form of BRST symmetric gaugeon formalism for the Higgs model. 
Our theory has seven gauge-fixing parameters $\alpha_i$ ($i=1,2,3$) 
and $\beta_j$ ($j= 1, 2, 3, 4$), some of which can be shifted by the 
$q$-number gauge transformation. The $q$-number gauge transformation commutes 
with the BRST transformation. As a result, the BRST charge is invariant, 
$\hat Q_\mathrm{B} = Q_\mathrm{B}$ and thus the physical subspace 
$\mathcal V_\mathrm{phys} = \ker Q_\mathrm{B}$ is also gauge invariant.

As a special choice of the gauge-fixing parameters ($\alpha_1=\varepsilon a^2, 
\alpha_2=\varepsilon a, \alpha_3=\varepsilon; \beta_j=0$), our theory includes 
the BRST symmetric version of the gaugeon formalism for Lorenz gauge by
Yokoyama and Kubo (\ref{L:YK}). 

Other choices of the parameters $\alpha$ and $\beta$ lead us to 
the theories of R$_\xi$-like gauges. 
Especially, by choosing (\ref{parameter_choice:Yao}), we get 
the gaugeon formalism for the Yao's R$_\xi$ gauge (\ref{L:Yao}), 
where one of the two gauge-fixing parameters, $\eta$, can be shifted by the 
$q$-number gauge transformation. In particular, the $q$-number 
gauge transformation can shift the $\eta$ 
to be equal to $\xi$, where the theory becomes equivalent to 
the R$_\xi$ gauge of Fujikawa, Lee and Sanda (\ref{L:FLS}).  
In any case of these R$_\xi$-like gauges, the $\xi$-parameter 
is shown to be invariant under the $q$-number gauge transformation. 

The invariance of the $\xi$-parameter under the $q$-number 
gauge transformation might be understood by the following
arguments. 
The propagator 
$\langle A_\mu A_\nu\rangle$ in the R$_\xi$ gauge of Fujikawa, Lee and Sanda 
is given by 
\begin{equation}
   \langle A_\mu A_\nu \rangle 
      = \frac{1}{p^2 - M^2}
        \left( - g_{\mu\nu} + \frac{p_\mu p_\nu}{M^2}
        \right)
        - \frac{p_\mu p_\nu}{M^2\,(p^2-\xi^{-1}M^2)}. 
\end{equation}
Now assume that there might exist a $q$-number gauge transformation 
$\hat A_\mu = A_\mu + \tau \partial _\mu \Lambda$, under which 
the $\xi$-parameter would transform into $\hat \xi=\xi + \Delta \xi$ ($\neq \xi$). 
Then, 
\begin{align}
   \langle \hat A_\mu \hat A_\nu \rangle 
      - \langle A_\mu A_\nu \rangle 
   &=\tau \left(
                 \langle A_\mu \partial_\nu \Lambda \rangle 
                + \langle \partial_\mu \Lambda A_\nu \rangle 
          \right)
     + \tau ^2 
                \langle \partial_\mu \Lambda \, \partial_\nu \Lambda \rangle 
  \notag 
  \\
  &=
  -\frac{p_\mu p_\nu}{M^2}
       \left[ 
             \frac{1}{p^2 - (\xi+\Delta \xi)^{-1}M^2}
           -
             \frac{1}{p^2 - \xi^{-1}M^2}
        \right]
  \notag
  \\
  &=
  -\frac{p_\mu p_\nu}{M^2}\,\frac{1}{p^2 - \xi^{-1}M^2}
     \left[ \left(
           1 -   \frac{\Delta M^2}{p^2 - \xi^{-1}M^2}
            \right)^{-1}
          -1
    \right]
  \notag
  \\
  &=
  -\frac{p_\mu p_\nu}{M^2}
       \left[ 
            \frac{\Delta M^2}{(p^2 - \xi^{-1}M^2)^2}
            + \frac{\Delta M^4}{(p^2 - \xi^{-1}M^2)^3}
            + \cdots
        \right], 
\end{align}
where $\Delta M^2= (\xi+\Delta \xi)^{-1}M^2 - \xi^{-1}M^2$.  
This 
shows that the field $\Lambda$ should include 
dipole modes, 
tripole modes, quadrapole modes, $\dots$,  
and any other higher-pole modes. 
The $Y$ field of 
our theory, however,  does not 
satisfies this condition: $Y$ includes at most quadrapole 
modes (see, for example, (\ref{propagator:PiPi})).
Thus we may infer  that 
the gaugeon formalism with  $\xi$-parameter which might be shifted 
by the $q$-number gauge transformation, if exists,  would require 
infinite series 
of multi-pole fields ($n$-pole fields with $n=2, 3, 4, \cdots$).  

We have seen in the section 5-1 that the Fock space of Yao's  R$_\xi$ gauge 
is embedded in the total Fock space of our theory
(if we choose  $\alpha_3=0$, $\beta=\xi^{-1}\boldsymbol 1$). 
In the arguments, the four 
 BRST-like charges (\ref{Q_BKO})-(\ref{Q_BY'})
(or equivalently, 
$Q_\mathrm{B(KO)}$,
$Q_\mathrm{B(Y)}$, 
$\hat Q_\mathrm{B(KO)}$ and 
$\hat Q_\mathrm{B(Y)}$)
play an essential role.
Similar arguments on the gauge structure of the Fock spaces 
are applicable to the theory in  Lorenz gauge  of Yokoyama and Kubo ($\beta=0$), 
since the four BRST-like charges (\ref{Q_BKO})-(\ref{Q_BY'}) also exist in
this gauge.
Here we shall consider 
the number of the conserved BRST-like charges in general case, 
and in what case the number becomes four. 
A BRST-like current may be expressed as 
\begin{equation}
     J^\mu_R= \mathcal B^\mathrm T R \overleftrightarrow \partial_\mu  \mathcal C,
\end{equation}
where 
$R$ is  a real and constant $2\times 2$ matrix.  
By using the field equations 
(\ref{eq:Box_calB}) and (\ref{eq:Box_calC}) we can evaluate the 
divergence of the current: 
\begin{align}
   \partial_\mu J^\mu_R 
     &= \mathcal B^\mathrm{T} (- \overleftarrow{\square}) R \mathcal C
       + \mathcal B^\mathrm{T} R \square \mathcal C
    \notag
    \\
    &= \mathcal B^\mathrm{T} M^2 ( \beta R - R \beta) \mathcal C.
\end{align}
Thus, the current $J_R^\mu$ is conserved if and only if 
\begin{equation}
          [R, \beta] =0. 
            \label{eq:[R,b]}
\end{equation}
The number of the independent matrices $R$ satisfying (\ref{eq:[R,b]}) 
is just the number of the conserved BRST-like charges. 
If $\beta$ is not proportional to the unit matrix, two types of the 
matrix 
$R=\mathrm{const.}\times \boldsymbol 1$ and 
$R=\mathrm{const.}\times \beta$ commute with $\beta$, thus in this case 
the number of the  conserved BRST-like charges is two.
On the other hand, if $\beta=\mathrm{const.}\times \boldsymbol 1$, 
an arbitrary matrix $R$ 
commutes with the $\beta$, thus 
there exist four independent conserved  BRST-like charges in this case. 
This is nothing but 
the case of Yao's gauge ($\beta=\xi^{-1}\boldsymbol 1$) and 
the Lorenz gauge of Yokoyama and Kubo 
($\beta=0$); 
no other case ensures the existence of four 
conserved currents. 


\section*{Acknowledgements}
We would like to thank Professor Imachi for his encouragement.

%

%
\end{document}